\documentclass[aps,prd,twocolumn,showpacs,nofootinbib]{revtex4-1}

\usepackage{amssymb} \usepackage{amsmath} \usepackage{graphicx}
\usepackage{epsfig,latexsym}
\usepackage{mathtools}

\RequirePackage{xspace} \allowdisplaybreaks

\begin{document}

\def\bef{\begin{figure}}
\def\eef{\end{figure}}

\newcommand{\nl}{\nonumber\\}

\newcommand{\ans}{ansatz }
\newcommand{\be}[1]{\begin{equation}\label{#1}}
\newcommand{\beq}{\begin{equation}}
\newcommand{\ee}{\end{equation}}
\newcommand{\beqn}[1]{\begin{eqnarray}\label{#1}}
\newcommand{\eeqn}{\end{eqnarray}}
\newcommand{\bd}{\begin{displaymath}}
\newcommand{\ed}{\end{displaymath}}
\newcommand{\mat}[4]{\left(\begin{array}{cc}{#1}&{#2}\\{#3}&{#4}
\end{array}\right)}
\newcommand{\matr}[9]{\left(\begin{array}{ccc}{#1}&{#2}&{#3}\\
{#4}&{#5}&{#6}\\{#7}&{#8}&{#9}\end{array}\right)}
\newcommand{\matrr}[6]{\left(\begin{array}{cc}{#1}&{#2}\\
{#3}&{#4}\\{#5}&{#6}\end{array}\right)}
\newcommand{\cvb}[3]{#1^{#2}_{#3}}
\def\lsim{\raise0.3ex\hbox{$\;<$\kern-0.75em\raise-1.1ex
e\hbox{$\sim\;$}}}
\def\gsim{\raise0.3ex\hbox{$\;>$\kern-0.75em\raise-1.1ex
\hbox{$\sim\;$}}}
\def\abs#1{\left| #1\right|}
\def\simlt{\mathrel{\lower2.5pt\vbox{\lineskip=0pt\baselineskip=0pt
           \hbox{$<$}\hbox{$\sim$}}}}
\def\simgt{\mathrel{\lower2.5pt\vbox{\lineskip=0pt\baselineskip=0pt
           \hbox{$>$}\hbox{$\sim$}}}}
\def\unity{{\hbox{1\kern-.8mm l}}}
\newcommand{\eps}{\varepsilon}
\def\ep{\epsilon}
\def\ga{\gamma}
\def\Ga{\Gamma}
\def\om{\omega}
\def\omp{{\omega^\prime}}
\def\Om{\Omega}
\def\la{\lambda}
\def\La{\Lambda}
\def\al{\alpha}
\newcommand{\ov}{\overline}
\renewcommand{\to}{\rightarrow}
\renewcommand{\vec}[1]{\mathbf{#1}}
\newcommand{\vect}[1]{\mbox{\boldmath$#1$}}
\def\tm{{\widetilde{m}}}
\def\mcirc{{\stackrel{o}{m}}}
\newcommand{\Dm}{\Delta m}
\newcommand{\dm}{\varepsilon}
\newcommand{\tanb}{\tan\beta}
\newcommand{\nbar}{\tilde{n}}
\newcommand\PM[1]{\begin{pmatrix}#1\end{pmatrix}}
\newcommand{\up}{\uparrow}
\newcommand{\down}{\downarrow}
\def\omE{\omega_{\rm Ter}}

%

\newcommand{\Dsusy}{{susy \hspace{-9.4pt} \slash}\;}
\newcommand{\DCP}{{CP \hspace{-7.4pt} \slash}\;}
\newcommand{\mc}{\mathcal}
\newcommand{\gr}{\mathbf}
\renewcommand{\to}{\rightarrow}
\newcommand{\gtc}{\mathfrak}
\newcommand{\wh}{\widehat}
\newcommand{\br}{\langle}
\newcommand{\kt}{\rangle}


\def\lsim{\mathrel{\mathop  {\hbox{\lower0.5ex\hbox{$\sim$}
\kern-0.8em\lower-0.7ex\hbox{$<$}}}}}
\def\gsim{\mathrel{\mathop  {\hbox{\lower0.5ex\hbox{$\sim$}
\kern-0.8em\lower-0.7ex\hbox{$>$}}}}}

\def\nn{\\  \nonumber}
\def\de{\partial}
\def\brf{{\mathbf f}}
\def\bbf{\bar{\bf f}}
\def\bF{{\bf F}}
\def\bbF{\bar{\bf F}}
\def\bA{{\mathbf A}}
\def\bB{{\mathbf B}}
\def\bG{{\mathbf G}}
\def\bI{{\mathbf I}}
\def\bM{{\mathbf M}}
\def\bY{{\mathbf Y}}
\def\bX{{\mathbf X}}
\def\bS{{\mathbf S}}
\def\bb{{\mathbf b}}
\def\bh{{\mathbf h}}
\def\bg{{\mathbf g}}
\def\bla{{\mathbf \la}}
\def\bmu{\mathbf m }
\def\by{{\mathbf y}}
\def\bmu{\mbox{\boldmath $\mu$} }
\def\bsig{\mbox{\boldmath $\sigma$} }
\def\bunity{{\mathbf 1}}
\def\cA{{\cal A}}
\def\cB{{\cal B}}
\def\cC{{\cal C}}
\def\cD{{\cal D}}
\def\cF{{\cal F}}
\def\cG{{\cal G}}
\def\cH{{\cal H}}
\def\cI{{\cal I}}
\def\cL{{\cal L}}
\def\cN{{\cal N}}
\def\cM{{\cal M}}
\def\cO{{\cal O}}
\def\cR{{\cal R}}
\def\cS{{\cal S}}
\def\cT{{\cal T}}
\def\eV{{\rm eV}}

%

\title{$\infty-\infty$: vacuum energy and  virtual black-holes}

\author{Andrea Addazi$^1$}\email{andrea.addazi@infn.lngs.it}
\affiliation{$^1$ Dipartimento di Fisica,
 Universit\`a di L'Aquila, 67010 Coppito AQ and
LNGS, Laboratori Nazionali del Gran Sasso, 67010 Assergi AQ, Italy}

\begin{abstract}

We discuss other contributions to the vacuum energy of quantum field theories 
and quantum gravity, which have not been considered in literature. 
As is well known, the presence of virtual particles in vacuum provides 
the so famous and puzzling contributions
to the vacuum energy. 
As is well known, these mainly
come from loop integrations over the four-momenta space. 
However,  we argue that these also imply the presence 
of a mass density of virtual particles in every volume cell of space-time. 
The most important contribution comes from 
 quantum gravity $S^{2}\times S^{2}$ bubbles, corresponding to virtual 
black hole pairs. 
The presence of virtual masses could lead to another paradox: the space-time itself would have an intrinsic virtual mass density contribution
leading to a disastrous contraction 
- as is known, no negative masses exist in general relativity. 
We dub this effect {\it the cosmological problem of second type}:
if not other counter-terms existed, 
the vacuum energy would be inevitably destabilized by virtual-mass contributions. 
It would be conceivable that the cosmological problem of second type could 
solve the first one. 
Virtual masses renormalize the vacuum energy 
to an unpredicted parameter, as in the  renormalization procedure
of the Standard Model charges. In the limit of $M_{Pl}\rightarrow \infty$ (Pauli-Villars limit), virtual black holes have 
a mass density 
providing an infinite  counter-term to the vacuum energy divergent contribution $M_{Pl} \rightarrow \infty$ (assuming $M_{UV}=M_{Pl}$).  
Therefore, in the same Schwinger-Feynman-Tomonaga attitude, the problem of a divergent vacuum energy could
be analogous to the {\it put-by-hand} procedure used for Standard Model parameters.

\end{abstract}

\maketitle
\section{Introduction}

As is well known, the Standard Model of particles predicts the existence of a zero point energy density $\rho_{vacuum} \rightarrow \infty$.
In an effective field theory approach, this is
interpreted as 
$$\rho_{vacuum}\sim -(n_{B}-\zeta(E)n_{F})M_{UV}^{4}$$
where $n_{B}$ and $n_{F}$ are the numbers of bosons and fermions,
$\zeta(E)$ is a coupling ratio running with energy, $M_{UV}$ is the UV Pauli-Villars cutoff scale.
In fact, bubble diagrams of matter, gauge sector and Higgs sector will badly generate quartic divergences
at the leading order plus subdominant divergences. 
 Such a vacuum energy would lead to a sudden super-inflation of space-time 
 in a Planckian (cosmological frame) time. 
 On the other hand, a very tiny dark energy density is observed, i.e. around $10^{-122}\, \rm M_{Pl}^{4}$. 
 This is the well known cosmological problem: even starting with a zero or very small Einstein's cosmological constant, 
 this will be destabilized by quantum corrections, up to $\Lambda \rightarrow M_{UV}^{2}$.
In (SM) $+$ (quantum gravity), the UV cutoff is the Planck scale $M_{UV}=M_{Pl}\simeq 10^{19}\, {\rm GeV}$
leading to a vacuum energy density $10^{122}\rho^{exp}_{\Lambda}$, where $\rho_{\Lambda}^{exp}$ is the energy density associated 
to the experimental cosmological constant. 
 
In this paper, we 
will discuss another aspect of the cosmological problem
related to a second paradox of quantum field theories and quantum gravity. 
We argue that quantum field theories inevitably lead to the presence of virtual particles 
in a volume cell of space-time.  
This implies  that, in every volume cell, virtual particles contribute to the energy density
with their own masses. In other words, the expectation value of the energy-momentum tensor 
gets a contribution from virtual masses floating in vacuum. 
This effect is analogous to virtual electric charges in electrodynamics, 
screening bare electric charges. 
However, the difference is that negative and positive electric charges contribute as 
$\langle 0|Q|0 \rangle=0$ while negative masses cannot exist in general relativity. 
Even if this effect is an obvious consequence of quantum field theory coupled to gravity, 
it was never discussed  as 
a relevant contribution to the vacuum energy density in any papers or reviews in literature
(See for example some reviews on these subjects
\cite{Weinberg:1988cp,Dolgov:1997za,Straumann:1999ia,Sahni:1999gb,Weinberg:2000yb,Carroll:2000fy,Rugh:2000ji,Padmanabhan:2002ji,Yokoyama:2003ii,Polchinski:2006gy,Martin:2012bt}).

The most important contribution comes from 
{\it Virtual Black Holes} (VBHs). 
Heisenberg's indetermination principle will inevitably lead to the conclusion that, 
localizing the position $\Delta x \rightarrow 0$,  the curvature fluctuations will diverge
$\Delta R \rightarrow \infty$. This leads to the pair creation of virtual black holes in vacuum. 
The quantum gravity scale imposes a bound on the Schwarzschild micro-black hole
mass of $M_{BH}\geq \sqrt{\pi} M_{Pl}$. 
Such a bound can be obtained as a solution of
$\lambda_{c}=r_{S}$, where $\lambda_{c}$ is the universal definition of Compton wave-lenght 
$\lambda_{c}=2\pi/M$ while $r_{S}=2G_{N}M$ is the Schwarzschild radius. 
As pointed out by S. Hawking, a virtual black hole bubble is topologically equivalent to 
$S^{2}\times S^{2}$, obtained by gluing two Eguchi-Hanson metric with opposite directions \cite{Hawking:1995ag}.
The Heisenberg's principle inevitably suggests the average presence of virtual micro black holes
with a minimal mass $\sqrt{\pi}M_{Pl}$ in a Planck volume $l_{Pl}^{3}$. 
Other larger mass virtual black holes clearly are possible, but the saturation density limit must be $\sqrt{\pi}M_{Pl}/l_{Pl}^{3}$, 
i.e. the black hole density is cut-off by the Planck density scale.
For instance a micro-black hole with $2\sqrt{\pi}M_{Pl}$ will occupy $2l_{Pl}^{3}$,
one with $3\sqrt{\pi}M_{Pl}$ will occupy $3l_{Pl}^{3}$ and so on.
As shown by S. Hawking, the quantum gravity ground state 
minimizing the semiclassical action corresponds 
to approximately one Planck mass micro-black hole 
for Planck volume \cite{Hawking2}. 
So that, a {\it space-time foam} with a leading topology $(S^{2})^{2N}$ with $N>>1$ was suggested
\footnote{But the same S. Hawking has not pointed out the possible relevance of the contribution of virtual BH masses to the renormalization of the vacuum energy. }
\footnote{More precisely, also the presence of $K_{3},\bar{K}_{3}$ are possible, while $CP_{2},\bar{CP}_{2}$
are not possible for a spin-structured metric gravity. }.
This leads to the catastrophic conclusion that space-time has an enormous intrinsic energy density of $\rho_{vacuum}\sim M_{Pl}^{4}$
even without on-shell matter sources. In fact if a space-time has an intrinsic energy density, this will source its 
curvature to $\langle R \rangle \rightarrow l_{Pl}^{-2}$. 

Of course, the BH mass bound can be deformed for electrically charged and/or rotating virtual black holes. 
More precisely, for non-extremal Kerr black holes with $M>J$, 
the mass bound is obtained solving the equation $\lambda_{c}=r_{outer} $, where 
$r_{outer}=\frac{1}{2}\left(r_{S}+\sqrt{r_{s}^{2}-4\alpha^{2}\cos^{2}\theta}\right)$, 
where $\alpha=J/M$ and $J$ the Spin parameter,
where $\theta$ parametrizes the angular variation of the radius in the oblate spheroidal horizon.

In the next subsection, we will provide a complete introduction to 
the classical cosmological problem 
and its completion with the second cosmological problem
introduced in this paper.

\subsection{Cosmological problems}

Let us consider the
Einstein-Hilbert lagrangian with a bare cosmological constant
and coupled with matter:
\be{SS}
S=\frac{1}{2\kappa}\int d^{4}x \sqrt{-g}(R-2\Lambda_{B})+S_{matter}[g_{\mu\nu},\Phi]
\ee
where $\kappa=8\pi G_{N}=1/M_{Pl}^{2}$
where $M_{Pl}$ is the reduced Planck mass.
In quantum field theory, 
the vacuum expectation value of the 
energy momentum tensor 
is not zero even without any matter sources: 
\be{Tmunu}
\langle 0|T_{\mu\nu}|0 \rangle=-\langle 0| \frac{2}{\sqrt{-g}}\frac{\delta W}{\delta g^{\mu\nu}} |0\rangle =-\rho_{vac}g_{\mu\nu}
\ee
where 
\be{W}
W=-i \log Z[0],\,\,\,Z[0]=\int \mathcal{D}{\Phi}e^{i S_{matter}[\Phi]}
\ee
and $\Phi$ are all SM particles and $Z$ is the full SM partition function, 
supposing to start the field quantizations from an almost flat background.

Eq.(\ref{Tmunu}) contributes to the Einstein's field equation 
as 
\be{con}
R_{\mu\nu}-\frac{1}{2}Rg_{\mu\nu}+\Lambda_{B}g_{\mu\nu}=\kappa T_{\mu\nu}^{matter}+\kappa \langle T_{\mu\nu} \rangle
\ee
In other words, the effective cosmological constant 
is 
\be{eff}
\Lambda_{eff}=\Lambda_{B}+\kappa \rho_{vac}
\ee
The SM complete vacuum energy is
\be{rho}
\rho_{vac}=\rho_{B}+\sum_{i}c_{i}\frac{m_{i}^{4}}{64\pi^{2}}\ln\left(\frac{m_{i}^{2}}{\mu^{2}} \right)+\rho_{vac}^{UV}+\rho_{vac}^{EW}+\rho_{vac}^{QCD}+...
\ee
$$\simeq \rho_{B}-2 \times 10^{8}\, {\rm GeV}^{4}++\rho_{vac}^{UV}+\rho_{vac}^{EW}+\rho_{vac}^{QCD}+...$$
where $\sum_{i}$ is done all over Standard Model particles
and in particular we use current data
$m_{H}\simeq 125\, {\rm GeV}$;
$c_{quarks}\simeq -4$, $m_{u}\simeq 2.3\, {\rm MeV}$,
$m_{d}\simeq 4.6\, {\rm MeV}$, $m_{s}\simeq 0.104\, {\rm GeV}$
$m_{c}\simeq 1.27\, {\rm GeV}$,
$m_{b}\simeq 4.2\, {\rm GeV}$, $m_{t}\simeq 171.2\, {\rm GeV}$;
 $m_{e}\simeq 0.511\, {\rm MeV}$, 
 $m_{\mu}\simeq 105\, {\rm MeV}$,
 $m_{\tau}\simeq 1.77\, {\rm MeV}$,
 (we ignore neutrinos),
 $M_{Z}\simeq 91\, {\rm GeV}$, 
 $m_{W_{\pm}} \simeq 80\, {\rm GeV}$,
 $c_{H}=1,c_{q}=-4,c_{l}=4,c_{Z}=c_{W}=3$;
 \be{UV}
 \rho_{vac}^{UV}=-\kappa \frac{1}{4}\sum_{F,B}\left[c^{B}(E)m_{B}^{4}(E)D^{B}(0)-c^{F}(E)m_{F}^{4}(E)D^{F}(0)\right]
 \ee
 where $D^{F,B}(0)$ are bubble diagrams $=\langle \Phi(0) \Phi(0) \rangle$
of all SM particles (F=fermions, B=bosons)  and we assume $M_{UV}=M_{Pl}$;
 \be{EW}
 \rho_{vac}^{EW}=-\frac{\sqrt{2}}{16}\frac{m_{H}^{2}}{G_{F}^{2}}\simeq -1.2\times 10^{8}\,{\rm GeV}^{4};
 \ee
 \be{QCD}
 \rho_{vac}^{QCD}\simeq (\langle \bar{q}q\rangle)^{4/3} \simeq 10^{-2}\, \rm GeV^{4};
 \ee
 Let us note that in Eq.\ref{UV}, $m_{B,F}^{4}D_{B,F}(0)\sim M_{UV}^{4}$ for $m_{B,F}\rightarrow 0$. 
 
 Now, let us include the contributions of virtual masses into Eqs.(\ref{UV},\ref{EW},\ref{QCD})
 \be{vmc1}
 \rho_{vac,TOT}^{VBH}= \rho_{vac}^{UV}+\sum_{VBH} c_{VBH}(E)  M_{Pl}^{4}
 \ee
 \be{vmc2}
 \rho_{vac,TOT}^{EW}= \rho_{vac}^{EW}+\sum_{SM}\frac{1}{8\pi^{3}} m_{SM}^{4}(E)
 \ee
 \be{vmc3}
\rho_{vac,TOT}^{QCD}= \rho_{vac}^{QCD}+\sum_{WL,G} \frac{1}{8\pi^{3}} m_{WL,G}^{4}(E)
 \ee
 where $\sum_{VBH}$ is on all virtual Black Hole species, 
$\sum_{SM}$ is performed on all SM particle masses, 
$\sum_{WL}$ is performed all over the
quark-antiquark Wilson lines (WL) and glueballs (G). 
For instance the mass of SM particle is averagely distributed 
in every associated Compton volume  $\lambda_{c}^{3}\sim m_{SM}^{-3}$
and so on. 
One can note that there is a coincidence in order of magnitude 
among various counter-terms. 
For example, the sum on all SM particles masses have approximately the same 
order of the vacuum expectation value but with an inverted sign. 
As well, QCD Wilson lines and Glueballs have a mass scale density of the same order $(1\div 10)$ of 
the QCD confinement energy density, but with again an inverted sign. 
Finally, the most important contributions coming from the Planck scale cutoff 
are from UV divergent loops and virtual black holes mass density, again expected 
with the same order of magnitude but inverted signs if $M_{UV}=M_{Pl}$. 
This alleviates the fine-tuning of the cosmological constant invoked in literature.
However, at this level of analysis is not possible to predict how much it can be alleviated, 
i.e. if the cancellation is down to namely $1/2$ or  $...$  or $10^{-50}$ or $...$ or $10^{-122}$ or $...$ or $<<10^{-122}$ of $O(1)M_{Pl}^{4}$. 
To demonstrate the exact cancellation among all these contributions 
seems impossible in framework of an effective quantum gravity theory. 
For instance, calculations of precise coefficients in Eq.(\ref{vmc1})
request an UV completed theory of quantum gravity 
because at $E\simeq M_{Pl}$ various radiative correction loops must contribute order $O(1)$
and they cannot be controlled in the effective quantum gravity framework. 
Another source of uncertainty comes from a possible unknown embedding of 
the SM in a larger group: clearly in this case the number of new particles 
as well as new introduced VEVs would contribute to the vacuum energy density.

\section{Path integral reformulation}

In this section, we will reformulate the cosmological problem in full SM $+$ quantum gravity path integral. 
All bubble diagrams contribute to the full partition function 
of Standard Model sector (or probably extensions) 
and quantum gravity: the partition function is nothing but 
an amplitude 
$Z=\langle vacuum| vacuum \rangle$.

The partition function of the Euclidean effective quantum gravity theory is the following path integral: 
\be{chi}
Z=\int \mathcal{D} g \mathcal{D}\psi \mathcal{D} \phi \mathcal{D} A_{\mu} e^{-I[g,\psi,\phi,A]} 
\ee
where $g$ is the Euclidean metric tensor and $\phi,A_{\mu},\psi$ are all SM scalars, bosons and fermions. 
The following argument can be generalized for a SM extension. 

In general the problem of quantum gravity is that this path integral has an UV divergence at 1-loop. 

In semiclassical limit 
\be{I}
I_{E}=-\int_{\Sigma}\sqrt{g}d^{4}x\left(\mathcal{L}_{m}+\frac{1}{16\pi}(R-2\Lambda)\right)
\ee
$$+\frac{1}{8\pi}\int_{\partial \Sigma}\sqrt{h}d^{3}x(K-K^{0})$$
where $\mathcal{L}_{m}$ is the matter Lagrangian and $G=c=1$. We will consider the case of $\Lambda=0$, i.e. the initial cosmological term is zero
$K$ is the trace of the curvature induced on the boundary $\partial \Sigma$
of the region $\Sigma$ considered, $h$ is the metric induced on the boundary $\partial \Sigma$, and 
$K^{0}$ is the trace of the induced curvature embedded in flat space.
The last term is a contribution from the boundary.
We consider infinitesimal perturbations of matter and metric
as $\phi=\phi_{0}+\delta \phi$,
$A=A^{0}+ \delta A$, (...)
and $g=g_{0}+\delta g$, 
so that 
\be{Is} I[\phi,A,...,g]=I[\phi_{0},A_{0},..g_{0}]+I_{2}[\delta \phi,\delta A,...\delta g]+\mbox{higher orders}\,,\ee
\be{Is1} I_{2}[\delta \phi,\delta A,..,\delta g]=I_{2}[\delta \phi,\delta A,...]+I_{2}[\delta g]\,.\ee
\be{logZ}
{\rm log} Z=- {\rm log} Z[\phi_{0},A_{0},...,g_{0}]
\ee
$$+{\rm log} \int \mathcal{D}\delta \phi \mathcal{D}\delta A(...) e^{-I_{2}[\delta \phi,\delta A,...]}+{\rm log}\int \mathcal{D}g e^{-I_{2}[\delta g]}$$

So that, according to Gibbons and Hawking, the Standard Model can be decoupled by gravitational fluctuations up to higher order corrections \cite{Euclidean1, Euclidean2}
in semiclassical limit. 
A conceptual problem of the decoupling limit comes from the fact the 
SM partition function is UV divergent and it is highly backreacting on the gravitational metric. 
At this point, the semiclassical limit seems not allowed. 
In other words, the full theory is not converging to a SM in a flat background in the limit of $E<<M_{Pl}$. 
The expectation value of the energy-tensor in vacuum has a non-zero component 
$$\langle vacuum| \tilde{T}_{00}| vacuum \rangle_{SM{-}integrals+EW+QCD}$$
But we have also to consider 
$$\langle vacuum| \tilde{T}_{00}| vacuum \rangle_{Virtual-Masses}$$
where 
$$\tilde{T}_{00}=T_{00}-\frac{1}{2}g_{00}T$$
where the main contribution of $\langle vacuum| \tilde{T}_{00}| vacuum \rangle_{Virtual-Masses}$ comes from 
$$\langle vacuum| \tilde{T}_{00}| vacuum \rangle_{S_{2}\times S_{2}}=c^{VBH}(E)M_{Pl}^{4}$$
corresponding to $\langle vacuum| I_{m}| vacuum \rangle \neq 0$.
A classification of possible quantum gravity bubbles can be done
on the basis of rigorous topological arguments \cite{Hawking:1995ag} . 
Quantum bubbles are classified by their topological invariants 
and in particular by their Eulero number and signature
\be{chi}
\chi=\nu+B_{2}^{+}+B_{2}^{-}=\frac{1}{128\pi^{2}}\int d^{4}x \sqrt{g} R_{abcd} R_{a'b'c'd'}\epsilon^{aba'b'} \epsilon^{cdc'd'}
\ee
\be{chi}
\tau=B_{2}^{+}-B_{2}^{-}=\frac{1}{96\pi^{2}}\int d^{4}x \sqrt{g} R_{abcd} R_{a'b'}^{cd}\epsilon^{aba'b'}
\ee
where $\nu=+2$ for compact manifolds and $\nu=+1$ for non-compact manifolds, 
and $B_{2}^{\pm}$ are the second Betti numbers of harmonic and anharmonic two 
forms respectively.
In the case of non-compact manifold boundary terms will contribute to the integral. 

All bubbles are topologically equivalent to only three possible classes: 
$S^{2}\times S^{2}, CP^{2}$ and $K^{3}$. 
Of course $CP^{2},K^{3}$ can be inversely oriented, 
 two possible anti-bubbles $\bar{CP}^{2},\bar{K3}$. 
 $S^{2}\times S^{2}$ has $(\chi,\tau)=(4,0)$,
$CP^{2}$ has $(3,1)$, $\bar{CP}^{2}$ has $(3,-1)$
$K3$ has $(24,16)$ and $\bar{K3}$ has $(24,-16)$. 
 In space-time with a spin structure, $CP^{2},\bar{CP}^{2}$ are not possible. 
 On the other hand $K3,\bar{K3}$ bubbles contribute to gravitational anomalies 
 and they can change helicity and chirality of particles. 
 But, their contribution in the path integral is expected to be subdominant:
Atiyah-Singer index theorem guarantees that they will have fermionic zero modes.
This implies that the only relevant diagrams are associated to $S^{2}\times S^{2}$, 
corresponding to virtual black holes pairs.

So that, the partition function can be fractioned in two parts as
\be{separation}
{\rm log} Z_{0}+{\rm log}Z_{\delta (SM), \delta g}
\ee
where $Z_{0}$ is referred to the partition function of SM in Euclidean flat background
and
\be{deltaSM}
Z_{\delta (SM), \delta g}=\int \mathcal{D}g \mathcal{D}\phi \mathcal{D}A_{\mu}\mathcal{D}\psi e^{-\langle\delta I_{m}\rangle+...}
\ee
where 
\be{coincidence}\langle \delta I_{m}\rangle=-\langle \delta I_{m} \rangle_{SM{-}integrals}+\langle \delta I_{m} \rangle_{Virtual{-}masses}+...
\ee
The vacuum energy will  decouple by gravitational $\sqrt{-g}$ inside the matter action $I_{m}$ if 
\be{deltaLm}
\langle \delta \mathcal{L}_{m} \rangle_{SM-Integral}=-\langle \delta \mathcal{L}_{m} \rangle_{Virtual-masses}
\ee
so that they decouple by gravitational factor $\sqrt{-g}$ inside the matter action $I_{m}$. 
Then a separation of (\ref{deltaSM}) is possible by virtue of the miraculous coincidence (\ref{coincidence}): 
  the partition function can be expressed as a product of separate Feynman's path integrals
  
  $$\left(\int \mathcal{D}\delta g \,e^{-I_{2}[\delta g]}\right)\left(\int \mathcal{D}\delta \phi ... \, e^{-I_{2}[\delta \phi,...]}\right)$$

More precisely, the virtual black hole action is 
\be{stf}
I_{1}\sim\frac{m_{BH}^{2}}{M_{Pl}^{2}}
\ee
and ignoring the interactions among $N$ black holes with the same mass (dilute gas approximation)
\be{stfn}
I_{N}\sim N\frac{m_{BH}^{2}}{M_{Pl}^{2}}
\ee
 we can evaluate the partition function of $N$ identical BH in a box of volume $V$
as
\be{Z}
Z_{0,BH}\sim \int_{0}^{\infty}dm_{BH} \sum_{N}\frac{1}{N!}\left(\frac{V}{l_{Pl}^{3}} \right)^{N}e^{-4\pi N\frac{m_{BH}^{2}}{M_{Pl}^{2}}}
\ee
This partition function is minimized by \cite{Hawking2}
\be{nBH}
\langle n_{BH} \rangle \sim 1,\,\,\,\langle m_{BH}^{2} \rangle \sim M_{Pl}^{2}
\ee
coinciding with its saddle point \footnote{
The integral on all BH masses is expected to be an approximation of a discrete
sum, i.e. BH masses are expected to be disposed in a discrete spectrum 
from arguments on the BH entropy \cite{Dvali:2011nh}. }. 
This implies that the spacetime must be repleted with Planck mass black holes, with a density of order one for Planckian volume.
This contributes to the complete partition function of SM and quantum gravity, as mentioned above. 
The problem is that virtual black holes are also gravitationally coupled with space-time 
and so that each others, i.e. the diluite gas approximation can be used only as a valid estimator 
of virtual black hole package. 
The curvature of space-time 
is maximized in modulus by (\ref{nBH}) if $M_{UV}\sim M_{Pl}$
and the main contribution is coming by virtual black holes gravitational attraction
$\langle V \rangle \sim G_{N} M_{Pl}^{2}/l_{Pl}\sim M_{Pl}$.

At this point we could be concern about possible non-perturbative quantum gravity corrections. 
For example, a black hole bubble can be corrected by infinite loops by gravitons, gauge bosons and so on.
We can expect that  any possible loop corrections to black-hole bubble
also correct SM bubbles, and the leading divergence is the same. 
Then, the limit of the space-time package of virtual Black holes cannot exceed 
$l_{Pl}^{4}$. But as stressed above, we have not the control of the running of couplings as mention above
and we can consider the problem only at the level of effective field theory. 
This is the no-go edge for our approach based on effective quantum gravity.

\subsection{$\Lambda\neq 0$ and real black hole pairs in de Sitter }

Now let us discuss the case of an initial cosmological constant $\Lambda \neq 0$. 
The presence of a bare cosmological constant can source the creation of a black hole pair in a de Sitter space. 
This process is mediated by a $S_{2}\times S_{2}$ gravitational instanton, generating an imaginary rate in the euclidean partition functional \cite{GP}. 
The topology of the gravitational instanton is exactly the same of virtual black hole bubbles, 
and 
the expansion energy can promote virtual black holes to a real pair. 
The nucleation rate is 
\be{nucleation}
\Gamma=Ae^{-(I_{S_{2}\times S_{2}}-I_{b.g.})}
\ee
where $A$ includes quantum corrections, $I_{S_{2}\times S_{2}}$ is the action associated to the gravitational instanton, 
$I_{b.g.}$ is the gravitational action of the background metric. 
As shown in \cite{GP}, the associated solution contains a negative mode. So that, the partition function becomes complex. 
This is the signal of an instability, leading to a spontaneous nucleation. 
The black holes are degenerate Schwarzschild-de Sitter black holes, or Nariai black holes, with Schwarzschild radius of $r_{H}=1/\sqrt{\Lambda}$ 
and tunneling rate $\Gamma\sim e^{-\pi G_{N}^{-1}\Lambda^{-1}}$. 
For $\Lambda<<G_{N}^{-1}\sim M_{Pl}^{2}$, the tunneling rate is exponentially suppressed. 
However, if $\Lambda=M_{Pl}^{2}$ the process cannot be calculated in semiclassical approximation 
and the system will disastrously nucleate into real black hole pairs with $100\%$ of probability. 
But if SM bubbles are cancelled by virtual black hole bubbles, they cannot source any instabilities. 

 Eq.(\ref{nucleation}) is related to partition functions of $S_{4}$ and $S_{2}\times S_{2}$ as 
\be{rate}
\Gamma=-\frac{1}{\pi}\sqrt{\frac{\Lambda}{3}}\frac{{\rm Im}Z[S_{2}\times S_{2}]}{Z[S_{4}]}
\ee
As shown in \cite{GP,Volkov:2000ih,Elizalde:1999dw,Nojiri:1998ph}, this path integral can be computed by integrating over 
the Fourier expansion coefficients of metric scalar, vector and tensor perturbations, 
obtaining an infinite product all over the eigenvalues. 
However, the conformal negative modes contributing to the integration are a finite set 
associated to the conformal operator 
 $\Delta_{0}=-3\nabla_{\mu}\nabla^{\mu}-4\Lambda$.
The final result is independent by the gauge-fixing choice
and the
Fadeev-Popov operator contains the zero-modes 
of the background isometries. 

 $\delta g_{\mu\nu}$ can be decomposed in scalar, vector and tensor modes: 
\be{decomposition}
\delta g_{\mu\nu}= \varphi_{\mu\nu}+\frac{1}{4}\delta g g_{\mu\nu}+\nabla_{\mu} \zeta_{\nu}+\nabla_{\nu} \zeta_{\mu}-\frac{1}{2}g_{\mu\nu}\nabla_{\sigma}\zeta^{\sigma}
\ee
under metric perturbations $g_{\mu\nu}+\delta g_{\mu\nu}$, the Einstein-Hilbert part of the action is perturbed as
\be{delta2}
\delta^{(2)} I_{\delta g}=\frac{1}{2}\langle \varphi^{\mu\nu},\Delta_{2}\varphi_{\mu\nu}\rangle-\frac{1}{16}\langle \tilde{h}, \Delta_{0}, \tilde{h}\rangle
\ee
where $\delta^{(2)}$ is the 2th order perturbation, 
$\tilde{h}=h-2\nabla_{\sigma}\zeta^{\sigma}$, 
\be{deltanabla2}
\Delta_{2}\varphi_{\mu\nu}=-\nabla_{\sigma}\nabla^{\sigma}\varphi_{\mu\nu}-2R_{\mu\nu\rho\sigma}\varphi^{\rho\sigma}
\ee
and the scalar product in a compact $\Lambda>0$ manifold is 
\begin{equation} \label{deltaa}
\langle \varphi_{\mu\nu}, \varphi^{\mu\nu}\rangle =\frac{1}{32\pi G_{N}}\int_{\Sigma}d^{4}x\sqrt{g}\varphi^{\mu\nu}\varphi_{\mu\nu}
\end{equation}
\be{deltab}
\langle \zeta_{\mu},\zeta^{\mu}\rangle =\frac{1}{32\pi G_{N}}\int_{\Sigma}d^{4}x\sqrt{g}\zeta_{\mu}\zeta^{\mu}
\ee
Let us consider the partition function perturbed at the 2th order: 
\be{contributionVari}
Z^{(2)}=\int \mathcal{D} g \mathcal{D} \phi \mathcal{D} A\mathcal{D}\psi e^{
I_{0}+
\frac{1}{2}\langle \varphi^{\mu\nu},\Delta_{2}\varphi_{\mu\nu}\rangle-\frac{1}{16}\langle \tilde{h}, \Delta_{0} \tilde{h}\rangle + \delta I_{m}+\delta I_{g.f.}}
\ee
where $\delta I_{m}$ contains SM and virtual black hole bubbles,
$\delta_{g.f}$ the a gauge fixing term
\be{gaugefixing}
\delta I_{g.f}=\gamma \langle \nabla_{\sigma} h^{\sigma}_{\rho}-\frac{1}{\alpha}\nabla_{\rho} h , \nabla^{\alpha} h_{\alpha}^{\rho}-\frac{1}{\alpha}\nabla^{\rho} h  \rangle
\ee
where $\gamma, \alpha$ are real constants. The gauge fixing term can be choose so that it not contains zero modes: $\gamma=1$, $\alpha=2$ \cite{Volkov:2000ih}.
Again, assuming $\langle \delta I_{m} \rangle=0$, the decoupling is safe. 

The bubble nucleation process is sourced by the classical action $I_{0}$ with $\Lambda>0$ 
causing negative modes the gravitational term (\ref{delta2}). 
On the other hand, we have shown that SM vacuum energy cannot source $O(1)$ nucleation instabilities
if completely screened by Virtual masses. 
At this point, the effective 2th order action is exactly the same solved in literature \cite{GP,Volkov:2000ih,Elizalde:1999dw,Nojiri:1998ph}, 
with exactly the same calculation procedure.

\section{Conclusions}
In this paper, we have shown how the cosmological constant is 
corrected by other counter-terms 
coming from the masses of virtual particles - still  not 
considered in the literature. 
The most important contribution comes from 
virtual black hole bubble diagrams. 
In order to distinguish this problem from the original stated cosmological problem, 
we define it {\it cosmological problem of the second type}.
The initial Einstein's cosmological constant is corrected combining integrated UV divergent SM bubble diagrams 
plus virtual masses. 
We have also discussed this problem in the path integral framework. 
In particular, we have discussed the cancellation mechanism with a null bare cosmological constant $\Lambda=0$
and with an initial small bare cosmological constant $\Lambda>0$. 
In doing this, we have tacitly assumed that CPT and unitarity are not violated by virtual black holes.
In other words, we have assumed that the information paradox would be not seriously interpreted 
as a violation of quantum mechanics principles. A possible decoherence effect by quantum gravity is assumed to be null. 
Of course, the information paradox is another different problem for an unknown microscopic theory of black holes
\footnote{We have suggested that semiclassical black holes could be reinterpreted as an effective geometry, composed of a large ensamble of horizonless naked singularities
(imagined smoothed at the planck lenght)
  \cite{Addazi:2015gna,Addazi:2015hpa,Addazi:2015bee}: the information paradox would be understood as a chaotization of the infalling information. 
  In this approach, a BH horizon does not exist and it is only an approximated geometry. 
  Paradoxical BHs are substituted by new geometric solutions dubbed {\it frizzyballs}, which recover all 
observable proprieties of astrophysical BH.
In frizzyballs, the no hair theorem is avoided and they carry informations in gravitational geometric hairs. }.

To conclude, 
the presence of UV divergent counter-terms to the usual cosmological vacuum energy 
implies a strong reduction of the cosmological fine-tuning suggested in literature. 
This also seems a relevant counter-argument against the usual disentanglement limit
of particle physics and quantum gravity, always applied in all quantum field theory calculations.
 Such a procedure seems 
justified in most of the scattering processes tested in laboratory.
But as shown above, such a procedure is not fully correct 
in the estimation of the vacuum energy density. 
For instance, the contribution of virtual black holes to the vacuum energy is 
not negligible as thought by most of theoretical physicists, 
but it is order $M_{Pl}^{4}$. 
Because of that, we hope that our arguments would be useful for a future clarification of the most serious unsolved problem of theoretical physics. 
The final dream is to demonstrate an exact cancellation among all terms in the framework of a UV completed theory of quantum gravity or a theory of everything.  
We think that this problem could be analogous to the old dream to calculate the electron mass from first principles
\footnote{ The cosmological problem has also a crucial importance in phenomenology:
if the SM vacuum energy density was zero despite of $10^{-122}M_{Pl}^{4}$, 
the cosmological term would be generated by an unknown mechanism
with possible predictions in phenomenology. 
For instance, we have suggested that the cosmological constant
could be generated by a dark strong condensate, 
also providing a natural candidate for dark matter
\cite{Addazi:2016sot,Addazi:2016nok}. 
Another interesting alternative is the reinterpretation of the vacuum energy density as 
generated by a Born-Infeld (BI) electrodynamic condensate. 
In this framework, we have shown how neutrino masses would be dynamically generated 
if neutrinos were coupled with the new BI vector boson \cite{Addazi:2016oob}. Finally, 
contributions of chiral symmetry violating bubbles as $K_{3},\bar{K}_{3}$ or more complicated 
space-time composed topologies can source a gravitational topological vacuum susceptibility 
 term $\langle \tilde{R}R\rangle \neq 0$ contributing to the vacuum energy and possibly generating a neutrino mass term.  
 This last possibility is connected to the recent suggestion of Ref.\cite{Dvali:2016uhn}. }.

\vspace{3cm}

{\large \bf Acknowledgments}
\vspace{3mm}

I would like to thank Massimo Bianchi, Salvatore Capozziello, Gia Dvali, Lena Funcke, Antonino Marciano and Sergei Odintsov 
for valuable discussions and remarks on these subjects. 
I also would like to thank
 Fudan University (Shanghai)
for the hospitality during the preparation of this paper. 
My work was supported in part by the MIUR research grant Theoretical Astroparticle Physics PRIN 2012CP-PYP7 and by SdC Progetto speciale Multiasse La Societ\'a della Conoscenza in Abruzzo PO FSE Abruzzo 2007-2013.

\end{document}